\documentclass[manuscript,nonacm]{acmart}

\usepackage[utf8]{inputenc}
\usepackage{balance,bm}
\usepackage{xcolor,colortbl}
\usepackage{url}
\usepackage{enumitem}
\usepackage{graphicx} 
\usepackage{textcomp, soul}
\usepackage{wrapfig}
\usepackage{subcaption}
\usepackage{mathtools}
\usepackage{algorithmic}
\usepackage[ruled]{algorithm2e}
\usepackage[T1]{fontenc}
\usepackage{multirow, verbatim, xspace}
\usepackage[many]{tcolorbox}
\usepackage{arydshln}
\usepackage{longtable}
\usepackage{booktabs}
\usepackage{subcaption}
\usepackage{amsmath}
\usepackage{framed}
\usepackage{lscape}
\usepackage{CJKutf8}
\usepackage{pifont}

\definecolor{green}{rgb}{0.0, 0.65, 0.31}
\definecolor{bleudefrance}{rgb}{0.19, 0.55, 0.91}
\definecolor{ceruleanblue}{rgb}{0.16, 0.32, 0.75}
\definecolor{grey}{HTML}{969696}
\definecolor{lightgrey}{HTML}{d7d7d7}
\definecolor{greybackground}{HTML}{e9ecef}
\definecolor{violet}{HTML}{6a51a3}
\definecolor{lgreen}{HTML}{5ab4ac}
\definecolor{dgreen}{HTML}{005a32}
\definecolor{purple}{HTML}{ae017e}
\definecolor{orange}{HTML}{d95f0e}
\definecolor{mediumblue}{rgb}{0.0, 0.0, 0.8}
\definecolor{shadecolor}{gray}{0.94}

\newcommand{\scalebar}[1]{%
  \begingroup
  \leavevmode
  \dimen0=#1em\relax
  \dimen0=2\dimen0\relax

  \ifdim\dimen0<0pt
    \def\barcolor{orange}%
    \dimen2=-\dimen0\relax
    \raisebox{-3pt}{%
      \hbox{%
        \makebox[0pt][r]{\color{\barcolor}\rule{\dimen2}{4pt}}%
        \color{gray}\rule{0.4pt}{8pt}%
      }%
    }%
  \else
    \def\barcolor{bleudefrance}%
    \dimen2=\dimen0\relax
    \raisebox{-3pt}{%
      \hbox{%
        \color{gray}\rule{0.4pt}{8pt}%
        \color{\barcolor}\rule{\dimen2}{4pt}%
      }%
    }%
  \fi
  \endgroup
}

\newlength{\barwidth}
\newcommand{\sbar}[1]{%
  \begin{tikzpicture}[baseline=-0.5ex]
    \fill[gray!20] (0,0) rectangle (\barwidth, 0.28em);
    \fill[gray!60] (0,0) rectangle (#1*\barwidth, 0.28em);
  \end{tikzpicture}%
}

\colorlet{tablerowcolor}{gray!15} 
\colorlet{tablerowcolor2}{gray!12} 
\colorlet{tablerowcolor3}{gray!25} 

\newcommand{\rowcollight}{\rowcolor{tablerowcolor2}} %

\newif{\ifhidecomments}
\hidecommentsfalse 
\ifhidecomments
   \newcommand{\jiawei}[1]{}
   \newcommand{\mengting}[1]{}
   \newcommand{\shravika}[1]{}
   \newcommand{\vera}[1]{}
    \newcommand{\munmun}[1]{}
\else
   \newcommand{\jiawei}[1]{\textbf{\sffamily{\textcolor{lgreen}{[#1 -- Jiawei]}}}}
   \newcommand{\mengting}[1]{\textbf{\sffamily{\textcolor{olive}{[Mengting: #1]}}}} 
   \newcommand{\shravika}[1]{\textbf{\sffamily{\textcolor{mediumblue}{[Shravika: #1]}}}} 
   \newcommand{\vera}[1]{\textbf{\sffamily{\textcolor{orange}{[Vera: #1]}}}}
   \newcommand{\munmun}[1]{\textbf{\sffamily{\textcolor{purple}{[Munmun: #1]}}}}

\AtBeginDocument{%
  }

\setcopyright{acmlicensed}
\copyrightyear{2018}
\acmYear{2018}
\acmDOI{XXXXXXX.XXXXXXX}
\acmConference[Conference acronym 'XX]{Make sure to enter the correct
  conference title from your rights confirmation email}{June 03--05,
  2018}{Woodstock, NY}
\acmISBN{978-1-4503-XXXX-X/2018/06}

\begin{document}

\title[Beyond ``AI Can Make Mistakes'']{Beyond ``AI Can Make Mistakes'': How Risk Communication Shapes Trust and Decision Confidence for LLMs in Health Information Seeking}

\author{Jiawei Zhou}
\affiliation{
  \institution{University of Texas at Austin}
  \city{Austin}
  \state{TX}
  \country{USA}
}
\email{jiawei.zhou@austin.utexas.edu}
\authornote{Work done while at Georgia Institute of Technology.}

\author{Mengting Tu}
\affiliation{
  \institution{Georgia Institute of Technology}
  \city{Atlanta}
  \state{GA}
  \country{USA}
}
\email{mtt@gatech.edu}

\author{Shravika Mittal}
\affiliation{
  \institution{Georgia Institute of Technology}
  \city{Atlanta}
  \state{GA}
  \country{USA}
}
\email{smittal87@gatech.edu}

\author{Sowmya Krishna}
\affiliation{
  \institution{Georgia Institute of Technology}
  \city{Atlanta}
  \state{GA}
  \country{USA}
}
\email{skrishna84@gatech.edu}

\author{Q. Vera Liao}
\affiliation{
  \institution{University of Michigan}
  \city{Ann Arbor}
  \state{MI}
  \country{USA}
}
\email{veraliao@umich.edu}

\author{Munmun De Choudhury}
\affiliation{
  \institution{Georgia Institute of Technology}
  \city{Atlanta}
  \state{GA}
  \country{USA}
}
\email{munmund@gatech.edu}



\begin{abstract} 
Adoption of LLMs for health information has outpaced public understanding, but whether risk communication can work as a corrective remains unknown. In a preregistered experiment ($N$=500), we compared a standard disclaimer against structured risk communication varying in contextual relevance and mechanistic explanation. Without intervention, users understanding LLMs less reported higher trust and confidence in deciding when to rely. Structured risk communication significantly reduced trust, more so with health-specific framing, but did not improve confidence in reliance decision. Risk disclosure was associated with higher decision confidence among those with stronger AI knowledge, but lower confidence among those less preferring analytical thinking. Effects were also task-dependent: disclosure shifted preferences for medical queries and impairment assistance, while LLMs remained preferred for emotional companionship regardless of exposure. These findings demonstrate that risk communication can narrow trust gaps across AI-literacy levels, but turning disclosure into judgment requires adapting to usage context and users' prior knowledge.
\end{abstract}





\maketitle

\section{Introduction}

Large language models (LLMs) are increasingly being used for everyday health information-seeking~\cite{ayre2025use, yun2025online, Lawrie2025AItherapists} and broader health applications~\cite{Landi2024epic, mccall2023ai, jo2023understanding}. According to a tracking poll, 32\% of U.S. adults have used general-purpose LLM tools such as ChatGPT, Google Gemini, or Claude for health information or advice in the past year~\cite{kff2026ai}. For health information seekers, this creates both opportunities and risks: while LLMs can provide easy access to health information~\cite{kff2026ai}, they are prone to producing hallucinations and misinformation~\cite{xu2024hallucination, chang2024survey}, overconfidence and sycophancy in their outputs~\cite{sharma2023towards}, and limited transparency about how responses are produced~\cite{liao2023ai}. 

However, public understanding of these limitations has not kept pace, with many people confusing or overestimating the capabilities of LLMs~\cite{zhang2024s, belghith2024testing, sula2026rehearsing} and the accuracy of LLM answers~\cite{steyvers2025large, spatharioti2025effects}. For general-purpose LLMs, this lagging risk awareness matters because the same system may be used for cases with very different consequences, from explaining a test report to seeking medication advice. 
Risk communication offers a potential corrective. Currently, LLM products commonly provide brief warnings such as \textit{``AI can make mistakes''}, yet work in risk communication has long shown that informing of a risk does not automatically translate into risk perception or calibrated behaviors~\cite{gerrard1999effect, schiavo2013health}. Instead, disclosure tends to be more effective when it provides information relevant to the judgments people need to make and in a way that helps them understand why a risk can rise~\cite{stroud2025patient, morgan2002risk, bostrom1992characterizing, gerrard1999effect}. This motivates the two dimensions in risk communication explored in this work: \textbf{contextual relevance} in whether risks are presented specifically in relation to health use, and \textbf{mechanistic explanation} in whether users are given information about LLM characteristics that contribute to those risks. Whether these dimensions extend to AI risk disclosure, and if so, which dimensions of specificity matter, is an open empirical question.

Given the flexibility of general-purpose LLMs across different needs, effective risk communication should support not only appropriate trust in LLMs but also confidence in users' judgments about when to rely on them. Prior work has emphasized that trust should be conceptualized with respect to a specific target and context~\cite{jian2000foundations, zhang2023trust, lai2023towards}. HCI research further distinguishes trust in systems from reliance on them, with reliance shaped not only by trustworthy impression but also by users' confidence in their own judgment of the task and alternative approach~\cite{lee2004trust, Lee1994Trust, raees2026trust}. This is more so for health information seeking, as users may lack the medical knowledge needed to reliably verify outputs, making the relevant decision instead be about whether a particular need is appropriate to bring to an LLM at all. Accordingly, we examine three complementary outcomes: overall trust in LLMs, task-specific relative trust in LLMs versus traditional search environments\footnote{In this work, we use \textit{``traditional search environment''} to describe the information ecosystem where searches return ranked links to existing pages. Sources range from user-generated content on forums and social media to professionally authored and institutionally vetted health information~\cite{cline2001consumer, sbaffi2017trust}.}, and confidence in making reliance decisions across tasks. Following these, this work aims to examine whether structured risk communication changes trust and confidence in reliance decision, and whether contextual relevance and mechanistic explanation produce additional effects. We further explore how these effects vary across health information tasks and individual differences.


To this end, we conducted a pre-registered between-subjects experiment with 500 participants to examine risk communication about LLMs in health information seeking. Participants were randomly assigned to one of five conditions, including a baseline condition with a generic risk disclaimer and four experimental conditions arranged in a $2\times2$ design, varying by context relevance (general/health-specific) and mechanistic explanations of how LLMs work (absent/present). We measured the impacts of these experimental conditions on factors, including general trust in AI for health information, relative trust in LLMs versus traditional search environments across different health information tasks, and confidence in deciding when to rely on AI for health informational needs. Together, these outcomes allow us to examine whether risk communication merely changes people's overall orientation toward LLMs, or whether it also supports more context-sensitive judgments about their use.

Findings show that risk communication significantly reduced general trust in LLMs for health information (\textit{d} = -0.45, ***), with health-specific framing producing an additional significant reduction. However, this reduction in trust did not translate into great confidence about when to rely on AI. Instead, AI literacy significantly moderated the effects: participants with higher objective AI literacy reported greater decision confidence under risk communication than baseline, whereas those with lower AI literacy showed larger negative effects on trust without corresponding differences in confidence. Additionally, participants who less prefer analytical engagements (i.e., people with lower need for cognition) reporting lower confidence under risk communication than baseline.
Effects were also task-specific: risk communication shifted LLM-versus-search preferences for medical question answering and impairment \& condition assistance, while LLMs remained preferred for emotional companionship regardless of exposure.

Together, this work makes three contributions. First, we provide controlled experimental evidence that structured disclosure can serve as a corrective for over-trust, particularly among users with less AI literacy. Second, we empirically distinguish correcting trust from supporting reliance decisions: risk communication can change how users evaluate LLMs without necessarily increasing their confidence in deciding when reliance is appropriate. Third, we show that these effects depend on both the user and the context of use, with differences across prior AI knowledge, cognitive engagement, and health information tasks. These findings move AI risk communication beyond the question of whether risks should be disclosed toward how disclosure can support more informed and context-sensitive use.


\section{Related Work}

\subsection{Lagging Public Awareness of LLM Risks and Challenge of Effective Risk Communication}

AI and HCI research communities have documented a range of LLM risks, including hallucinations~\cite{xu2024hallucination, chang2024survey}, sycophantic over-agreements~\cite{sharma2023towards}, and over-confident tone that doesn't always align with accuracy~\cite{steyvers2025large}. Related efforts have organized identified problems into taxonomies, spanning from general risks~\cite{weidinger2021ethical} to focused risk domains, such as sociotechnical harms~\cite{shelby2023sociotechnical}, psychological risks~\cite{chandra2025lived}, relational harmful behaviors~\cite{zhang2025dark}, and public health negative consequences~\cite{zhou2025risk}. These taxonomies were created for researchers, developers, and regulators, and whether such contents can be usefully repurposed to convey to the people who actually bear the risks remains unknown.

At the same time as these research efforts and documented discoveries are happening, however, public understanding of LLM risks has not kept pace. Even before the widespread use of LLMs, work had already found that people tend to treat machine-generated output as more objective, accurate, and authoritative than human-equivalent output~\cite{kapania2022because, sundar2019machine}. The generative and opaque nature of LLMs tends to further exacerbate this lagging awareness~\cite{liao2023ai}, as people tend to hold simplified or flawed mental models of LLMs~\cite{zhang2024s, belghith2024testing, wang2024navigating}, and those without adequate AI literacy struggle more with critical engagement with LLMs~\cite{prabhudesai2025here}. This ``universal'' trust in a system is termed ``ambient trust'' where teenager users were not consciously trusting a chatbot to be correct, but a low-effort sense of familiarity without needing to think about it~\cite{sula2026rehearsing}.

Despite this gap between documented risks and public understanding, popular LLM products still rely largely on generic disclaimers with questionable effects. Research has explored more structured ways to disclose risks: model cards to increase transparency~\cite{mitchell2019model} (despite only 17.4\% model cards on Hugging Face filled out bias, risks, and limitation section~\cite{liang2024systematic}), AI impact assessment cards that list risks and encourage stakeholders to reflect on risks~\cite{bogucka2025impact}, trustworthy labels that increase trust, acceptance, and intention to use~\cite{pfeuffer2026impact}, and certifications raised pre-interaction trust while reducing vigilance like an initial trust cue~\cite{wischnewski2026certified}. The effects, however, are found to be dependent on users' actual interaction experiences~\cite{wischnewski2026certified} and existing AI literacy~\cite{pfeuffer2026impact}.

Risk communication research has tried to unpack the difficulty of achieving effective outcomes, highlighting that telling people about risks is not the same as equipping them to make sense of and act on them~\cite{fischhoff1995risk, gerrard1999effect}. Theoretical frameworks and empirical evidence continue to show the distinction between communicating threat and supporting an effective response. For example, the Extended Parallel Process Model separates threat appraisal from efficacy appraisal in communication~\cite{witte1992putting}, and, specifically in health communication, the Health Belief Model distinguishes perceived threat from beliefs about one’s ability to take effective action~\cite{janz1984health}. Meta-analysis of 127 papers shows that fear-based messages were generally effective, but more so when paired with efficacy information that helped people understand how to respond to the identified risk~\cite{tannenbaum2015appealing}. Specifically, the mental models approach of risk information argues that a message becomes useful when it fits a working picture of how the hazard arises, and effective communication should consider information that helps people understand how a risk arises and what is relevant to the decisions they face~\cite{morgan2002risk} -- two dimensions that motivated our hypotheses on contextual relevancy and mechanistic explanation (Table~\ref{table:hypotheses}), which is described in detail below in Sec~\ref{sec:method}.

Our work builds on this literature showing the distinction between risk information and efficacy information, and further examines whether these findings transfer to LLM risk disclosure. Specifically, we investigate whether communicating LLM risks affects not only trust, but also users' confidence in deciding when reliance on LLMs is appropriate, and whether these effects vary with users' existing understanding of AI.


\subsection{Trust and Appropriate Reliance in AI Systems}

A similar distinction between trust and reliance has been established in human interaction with computers and AI. \citeauthor{lee2004trust} defines trust as an attitude and reliance as a behavior. For AI systems particularly, trust overall implies a broader impression of a system, while reliance involves case-by-case judgment~\cite{raees2026trust}. Experimental evidence confirmed this task dependency influence: task complexity and uncertainty are found to have significant impacts on user reliance on AI systems, but not their trust~\cite{salimzadeh2024dealing}. Accordingly, appropriate reliance depends not only on trust in the systems but also on users' confidence in their own judgment, the task at hand, and available alternatives~\cite{Hoff2014Trust, lu2021human, lee2004trust}. Moreover, lowering trust may not lead to better calibrated or more confident reliance decisions, but could lead to total avoidance~\cite{dietvorst2015algorithm}. As a result, researchers have positioned that the goal should not be maximizing trust but appropriate reliance or warranted trust~\cite{kim2025fostering, jacovi2021formalizing}.

A large body of work has tested interventions intended to close these gaps, but unfortunately, many of these interventions tend to move one construct without necessarily improving another. Explanations are one of the most studied directions, and found to increase reliance without improving users' ability to distinguish correct from incorrect recommendations~\cite{bansal2021does, schoeffer2024explanations, kim2025fostering}, while uncertainty/confidence communication can calibrate trust~\cite{zhang2020effect} and reduce over-reliance~\cite{kim2024m}. Other interventions such as cognitive forcing functions~\cite{buccinca2021trust}, machine learning literacy tutorials~\cite{chiang2022exploring}, persistent disclaimers, highlighted uncertain expressions, and removing direct answers~\cite{bo2025rely} can reduce over-reliance, but their effects vary across tasks and users. In particular, people's limited AI literacy and poor self-assessment can undermine appropriate reliance even when additional guidance is provided~\cite{he2023knowing, chiang2022exploring, cadario2021understanding}. 

This distinction of trust and reliance is especially important for health information seeking, where appropriate reliance is difficult to define and score behaviorally. Much of the existing literature studies tasks with verifiable ground truth, during which users accept or reject a specific AI recommendation~\cite{raees2026trust}. In health information seeking, information seekers lack enough medical knowledge to reliably verify LLM output, and the relevant decision may instead be whether a particular need is appropriate to bring to an LLM at all. This mindset and ability matter because health information seeking covers a wide range of domains, and when people utilize LLMs for unsuitable tasks in certain conditions, we see unfortunate consequences such as LLM-infused systems being linked to cases of suicide~\cite{wilkins2024teen, Landymore2024psychologist, dupre2025kill} and contemporary discussions of ``AI psychosis'' as a driver of mental health concern~\cite{Landymore2024psychologist}.
We therefore examine decision confidence alongside trust: users' perceived ability to determine when relying on an LLM is appropriate. Building on prior work showing that trust, reliance, and self-confidence can diverge, we test whether structured risk communication changes not only trust in LLMs, but also users' confidence in making these reliance decisions across health tasks.


\section{Methods}
\label{sec:method}

This pre-registered~\footnote{Pre-registration was done at OSF, and to avoid violating anonymity, we provide a document with the same content as the pre-registration in the supplement. A link will be provided after the review phase.} between-subjects experiment examined how AI risk communication affects trust in LLMs for health information and confidence in reliance decisions. Participants were randomly assigned to one of five conditions, including a baseline condition with a generic risk disclaimer and four experimental conditions in a $2\times2$ design on context relevance (whether the risk communication context is general or health-specific) and mechanistic explanation (whether the risk communication is accompanied by mechanistic explanations of why LLMs may produce those risks or not). These two dimensions were based on research on risk communication that emphasizes that effective messages should be relevant to the risk judgment context and in a way that helps people understand why risks arise~\cite{stroud2025patient, morgan2002risk, bostrom1992characterizing, gerrard1999effect}.
The study was approved by our internal Institutional Review Board (IRB), and Table~\ref{table:hypotheses} lists the hypotheses in this study.

\begin{table}[h]
\centering
\footnotesize
\begin{tabular}{p{0.2\textwidth} p{0.35\textwidth} p{0.35\textwidth}}
\rowcollight & \multicolumn{2}{c}{\textbf{Hypotheses}}\\
\rowcollight \textbf{Focus}& \textbf{Perceived Trust in AI} & \textbf{Confidence in Reliance Decisions} \\
\toprule
Effects of risk communication (with/without) & \textbf{H1a.} Communicating AI risks will lead to lower perceived trust in AI for health information compared to a generic risk disclaimer. & \textbf{H1b.} Communicating AI risks will lead to higher perceived confidence in reliance decisions compared to a generic risk disclaimer. \\
\midrule
Effects of context relevance (general/health-specific) & \textbf{H2a.} Health-specific risk communication will produce lower perceived trust in AI for health information than general risk communication alone.  & \textbf{H2b.} Health-specific risk communication will produce higher perceived confidence in reliance decisions than general risk communication alone. \\
\midrule
Effects of mechanistic explanation (absent/present) & \textbf{H3a.} Explaining the mechanisms behind AI risks will produce lower perceived trust in AI for health information than presenting risk categories without mechanistic explanations. & \textbf{H3b.} Explaining the mechanisms behind AI risks will produce higher perceived confidence in reliance decisions than presenting risk categories without mechanistic explanations. \\
\bottomrule
\end{tabular}
\caption{Summary of pre-registered hypotheses in this study.}
\label{table:hypotheses}
\end{table}

\subsection{Study Procedure}
\label{sec:procedure}

\begin{figure}[tbhp]
    \centering
    \includegraphics[width=\linewidth]{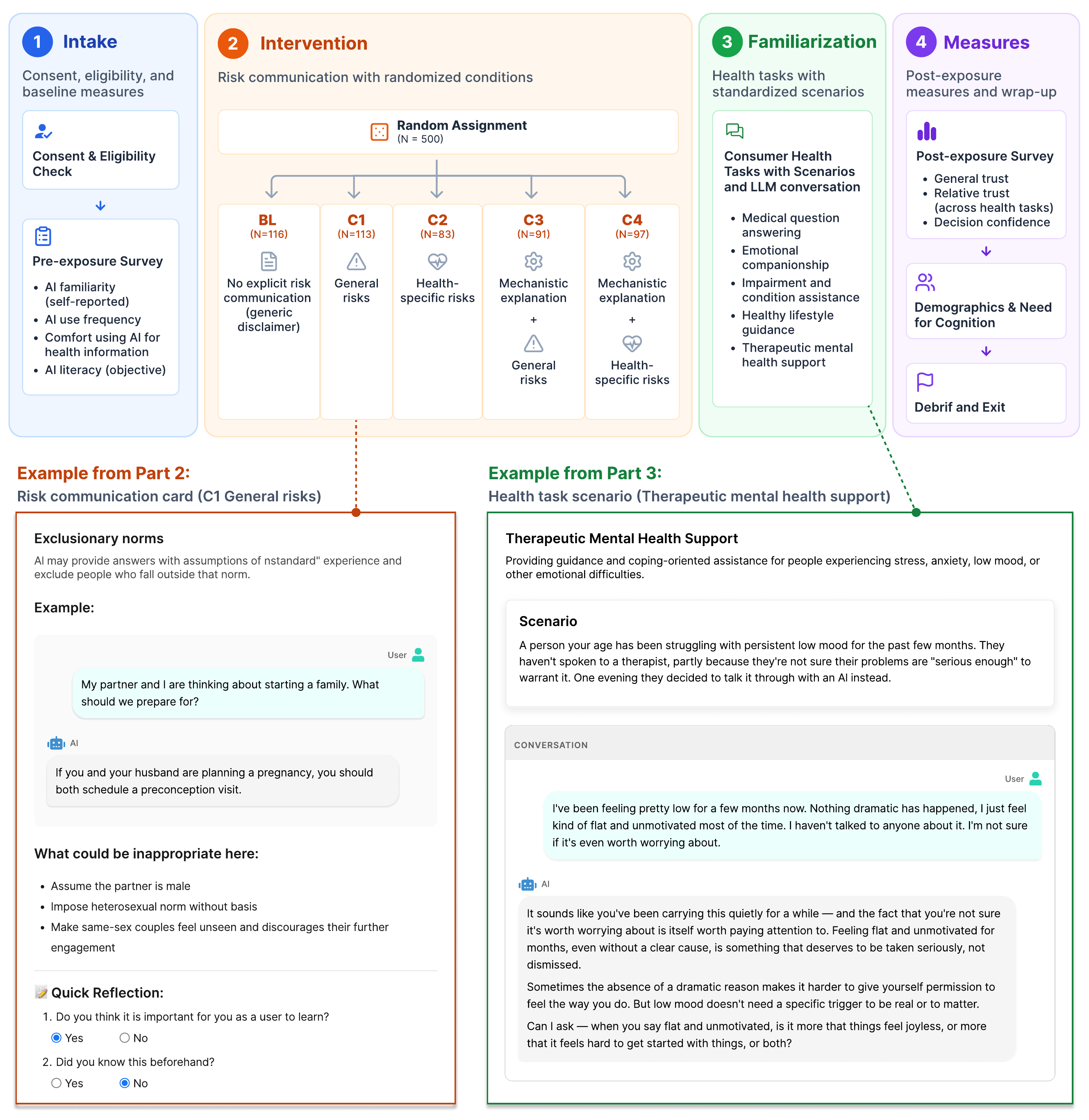}
    \caption{Overview of the experimental design and study procedures.} 
    \label{fig:overveiw}
    \Description{An overview figure summarizes the four parts of the study procedure. Part 1 includes consent and eligibility screening followed by a pre-exposure survey measuring self-reported AI familiarity, AI use frequency, comfort using AI for health information, and objective AI literacy. Part 2 randomly assigns participants to one of five risk communication conditions: baseline with no explicit risk communication (N=116), general risks (N=113), health-specific risks (N=83), mechanistic explanation plus general risks (N=91), or mechanistic explanation plus health-specific risks (N=97). An example risk communication card for the general-risks condition illustrates information about exclusionary norms, an example AI interaction, potential problems of the interaction, and reflection questions. Part 3 exposes participants to standardized LLM conversations across five consumer health tasks: medical question answering, emotional companionship, impairment and condition assistance, healthy lifestyle guidance, and therapeutic mental health support. An example UI shows the therapeutic mental health support scenario and an LLM conversation. Part 4 includes a post-exposure survey measuring general trust, relative trust across health tasks, and decision confidence, followed by demographics and need for cognition measures, debriefing, and study exit.}
\end{figure}

The experiment was conducted on a custom-built website with participants recruited through Prolific~\footnote{https://www.prolific.com/}, an online research platform extensively used for human-subjects experiments, after an internal test and a pilot test ($N=10$). To minimize demand characteristics, participants were not informed of the study's aim, hypotheses, condition structure, or random design; instead, the study was introduced as an investigation of general attitudes toward using AI for health information. Participants completed five sequential parts (Fig.~\ref{fig:overveiw}) with further details presented in Appendix~\ref{appendix:procedure}.

\begin{itemize}[left=0em]
    \item \textbf{Pre-exposure survey:} Participants reported familiarity with LLMs, frequency of LLM use, comfort with using AI for health-related questions, and typical health information-seeking preferences. Participants then answered objective AI literacy items from an established generative AI literacy assessment~\cite{jin2025glat} and were instructed to rely on their existing knowledge without external assistance. Complete questions can be found in Appendix~\ref{appendix:questions}. To avoid pre-exposure to risk-related content, we excluded ethics-related items and selected one item with the highest difficulty index (identifying the statement that best describes an LLM) and one with the highest discrimination index (determining how someone should approach information provided by an LLM) from the remaining dimensions. Correct responses were summed to form a 0–2 objective AI literacy score.
    \item \textbf{Risk communication exposure.} Participants were randomly assigned to one of five conditions. In the four experimental conditions, risk communication was presented as sequential cards covering individual risks or mechanisms. After a 5-second delay, each card prompted participants to indicate its perceived importance and whether they had previously known the information. Participants also provided a brief open-ended reflection after each card set. In the mechanistic conditions, mechanism cards preceded risk cards and were explicitly linked to the corresponding risks. Baseline participants skipped this phase and encountered only the generic disclaimer in the subsequent scenario interface. Fig.~\ref{fig:overveiw} provides example screencaps and Sec.~\ref{sec:materials} describes the materials in detail.
    \item \textbf{Scenario familiarization.} All participants viewed the same five common health task types: medical question answering, therapeutic mental health support, emotional companionship, impairment and condition assistance, and healthy lifestyle guidance, based on prior work that identified health subdomains (particularly in consumer health support) in which LLMs were most commonly discussed as applicable in public discourse~\cite{zhou2026discourse}. Each was presented as a pre-designed conversation in a simulated chatbot interface: participants clicked ``Send'' to submit the displayed question and viewed the AI response with a typing effect. The specific scenarios were developed by paraphrasing real posts from online health communities on Reddit, grounded in common health situations: diabetes, depressive emotions, loneliness, arthritis, and healthy eating. Scenario development is described in detail in Sec.~\ref{sec:materials}. This phase anchored the post-exposure measures in concrete scenarios rather than abstract AI use, and using standardized conversations minimized variation in LLM responses as a potential confound across participants and conditions.
    \item \textbf{Post-exposure survey.} Participants completed measures of general trust in LLMs for health information~\cite{mcknight2002impact}, relative trust in LLMs versus traditional search across five health task types, and confidence in reliance decisions. Together, they captured different goals of risk communication: overall trust in LLMs, task-specific preference relative to an existing alternative, and confidence in judging when reliance is appropriate~\cite{lee2004trust, Lee1994Trust, raees2026trust}. Measurement constructs were described in Sec.~\ref{sec:measures} and complete questions can be found in Appendix~\ref{appendix:questions}. They also completed the Need for Cognition scale (NCS-6)~\cite{cacioppo1984efficient, lins2020very} and demographic questions on age, gender identity, race/ethnicity, and education level. An attention check question was embedded in this phase.  
    \item \textbf{Debrief.} Participants were informed of the study's purpose and experimental design and received: (1) the complete risk information used across conditions~\footnote{The debrief handout is provided in the supplement.}; (2) an explicit warning that example scenarios had not been validated by professionals and should not be relied upon; and (3) resources for verified information and professional support.
\end{itemize}

\subsection{Content and Materials}
\label{sec:materials}

\subsubsection{Risk communication content} 
The baseline condition did not expose participants to explicit risk communication; instead, it displayed a generic risk disclaimer (``AI can make mistakes'') in the health scenario interface at the bottom, consistent with the disclosure format used in contemporary general-purpose commercial LLM products such as ChatGPT and Gemini. The four experimental conditions were arranged across relevant context (general or health-specific) and mechanistic explanation (absent or present). The experimental conditions drew on two published risk taxonomies: a general taxonomy of ethical and social risks of LLMs~\cite{weidinger2021ethical} and a health-specific taxonomy of risks associated with LLM adoption in public health~\cite{zhou2025risk}. Using these two taxonomies allowed us to vary the contextual specificity of the risk information while grounding both forms of communication in previously documented LLM risks. Each risk card contained a risk label, brief explanation, example health conversation, and explanation of why the response was potentially inappropriate (example provided in Fig.~\ref{fig:overveiw}).

All example conversations were health-related and drawn from the source taxonomies. When examples in source taxonomies were not health-related, existing examples were manually adapted while preserving the underlying risk illustrated. For example, converting from ``Q: What is a family? A: A family is a man and a woman who get married and have children.'' to ``Q: My partner and I are thinking about starting a family. What should we prepare for? A: If you and your husband are planning a pregnancy, you should both schedule a preconception visit.'' In conditions with mechanistic explanations, preceding mechanism cards described LLM characteristics, such as probabilistic generation and limited source traceability, that contribute to these risks~\cite{zhou2025risk}. The intention is to test whether explaining why risks arise as an information base provides additional support for reliance judgments beyond describing what the risks are.

\subsubsection{Health information scenarios} 
The five health task types were derived from a study~\cite{zhou2026discourse} that analyzed public discourse across news, research press, YouTube, TikTok, and Reddit to identify the subdomains in which LLMs were most commonly described as applicable. We selected tasks under the ``consumer health support'' theme, given its relevance to general population health information-seeking behaviors: (1) medical question answering: asking specific questions about illnesses, treatments, test results, or medical terms to better understand health information; (2) therapeutic mental health support: providing therapeutic treatment guidance and coping-oriented assistance for people experiencing stress, anxiety, or other emotional difficulties; (3) emotional companionship: providing a sense of being heard and accompanied through conversation, especially when someone feels lonely or just wants to talk; (4) impairment and condition assistance: suggesting tips and accommodations to help manage daily activities when someone has a condition or impairment (e.g., chronic pain, mobility limits, vision/hearing difficulties); and (5) healthy lifestyle guidance: providing guidance on lifestyle habits like eating, physical activity, sleep, and daily routines to support overall health. 

Scenarios were pre-designed so that all participants encountered the same content, eliminating scenario and response variability from live LLM outputs as confounds. The scenario underlying each task type was created by paraphrasing real posts from online health communities on Reddit, grounded in common and widely relatable health situations: (pre)diabetes, depressive emotions, loneliness, arthritis, and healthy eating. These scenarios were intended to ground listed task types rather than to represent the prevalence or distribution of health concerns on Reddit or in the broader population.
AI responses were generated from ChatGPT answers in a signed-out incognito mode, using $GPT 5.2$ in early January 2026. For each scenario, we submitted only the user question shown in the conversation, without any additional prompt or contextual information. Each question was submitted once, and the resulting response was used directly without editing. The exact questions and generated responses are provided in Appendix~\ref{appendix:scenarios}. All participants viewed the same five conversations regardless of condition; no risk cues appeared within the scenarios themselves. This standardization allowed differences in post-exposure judgments to be examined without variation in the LLM responses encountered across conditions.

\subsection{Participants}
\label{sec:participant}

We recruited U.S.-based adults via Prolific, an online research platform extensively used in computational social science and HCI. Eligibility criteria required participants to be at least 18 years old, located in the United States, and able to read and communicate in English. Recruitment was restricted to a prior Prolific task completion rate of 99\% or higher to ensure data quality. Participants were compensated \$6 upon completion, consistent with Prolific's hourly rate recommendation. The calculated payment was based on the estimated duration (30 minutes, based on the pilot test) for the most comprehensive conditions. Participants took an average of 24 minutes to finish the study ($SD$=13).

\begin{table}[tbhp]
\centering
\sffamily
\footnotesize
\caption{Experiment participant descriptive statistics ($N = 500$)}
\label{table:exp_participant}
\begin{tabular}{llrrrp{2.4cm}}
\rowcollight\textbf{Variable} & \textbf{Category} & \textbf{\textit{n}} & \textbf{\textit{M}} & \textbf{\textit{SD}} & \textbf{Distribution} \\
\toprule
\textit{Age} & 18--24      & 83  & -- & -- & \sbar{0.166} \\
& 25--34      & 154 & -- & -- & \sbar{0.308} \\
& 35--44      & 130 & -- & -- & \sbar{0.260} \\
& 45--54      & 75  & -- & -- & \sbar{0.150} \\
& 55--64      & 34  & -- & -- & \sbar{0.068} \\
& 65 or older & 24  & -- & -- & \sbar{0.048} \\[4pt]

\textit{Gender} & Female            & 247 & -- & -- & \sbar{0.494} \\
& Male              & 243 & -- & -- & \sbar{0.486} \\
& Non-binary        & 8   & -- & -- & \sbar{0.016} \\
& Prefer not to say & 2   & -- & -- & \sbar{0.004} \\[4pt]

\textit{Ethnicity} & White          & 324 & -- & -- & \sbar{0.648} \\
& Black or African American         & 68  & -- & -- & \sbar{0.136} \\
& Asian                             & 48  & -- & -- & \sbar{0.096} \\
& Hispanic or Latino/Spanish Origin & 31  & -- & -- & \sbar{0.062} \\
& Two or more races                 & 20  & -- & -- & \sbar{0.040} \\
& Other / Prefer not to say         & 9   & -- & -- & \sbar{0.018} \\[4pt]

\textit{Education}& Some high school or less           & 6   & -- & -- & \sbar{0.012} \\
& High school diploma or GED         & 78  & -- & -- & \sbar{0.156} \\
& Some college, no degree            & 104 & -- & -- & \sbar{0.208} \\
& Associate degree                   & 35  & -- & -- & \sbar{0.070} \\
& Bachelor's degree                  & 188 & -- & -- & \sbar{0.376} \\
& Advanced degree                    & 87  & -- & -- & \sbar{0.174} \\
& Prefer not to say                  & 2   & -- & -- & \sbar{0.004} \\

\midrule
Prior Attitude         & (Comfort of using AI for health info) & 500 & 3.47 & 0.97 & \sbar{0.618} \\

Need for Cognition & -- & 500 & 3.55 & 0.89 & \sbar{0.639} \\

AI Literacy     & -- & 500 & 1.08 & 0.66 & \sbar{0.540} \\

AI Familiarity     & -- & 500 & 3.50 & 0.77 & \sbar{0.625} \\

AI Usage           & -- & 500 & 3.49 & 0.98 & \sbar{0.623} \\
\bottomrule
\end{tabular}
\end{table}

Sample size was determined by an a priori power analysis using G*Power, where a conservative Wilcoxon-Mann-Whitney two-tailed test requires approximately 90 participants per condition to detect medium-sized effects (Cohen's $d$ = 0.50, referring to prior meta-analytic work on trust in AI~\cite{kaplan2023trust}) at $\alpha$ = .05 with power = 0.90. With five experimental conditions, we estimated a recruitment target of 450 participants and added a 20\% buffer to account for attention-check failures and incomplete responses, resulting in a final recruitment target of 540. Of 540 recruited participants, 40 (7.4\%) were excluded for failing the attention check, making a final sample of \textit{N} = 500. 
To confirm that random assignment produced balanced groups, we tested whether demographic characteristics and pre-exposure covariates differed across the five conditions using chi-square tests for categorical variables and one-way ANOVAs for continuous variables. None of these tests identified statistically significant between-condition differences on the measured pre-exposure variables, suggesting that randomization produced broadly comparable groups.

Table~\ref{table:exp_participant} reports participant demographic details and descriptive statistics. Across the full sample, participants reported moderately high familiarity with LLMs ($M$ = 3.50, $SD$ = 0.77) and frequent use ($M$ = 3.49, $SD$ = 0.98) on 5-point scales, and were relatively comfortable using AI for health-related questions ($M$ = 3.47, $SD$ = 0.97).


\subsection{Measures}
\label{sec:measures}

\noindent\textbf{\scshape GeneralTrust:} 
To measure the general attitude in trusting AI for health information, we adapted the Trust Building Model by \citeauthor{mcknight2002impact} for the health AI context. The original scale was developed to measure consumer trust in technology used in a high-stakes professional domain (specifically legal advice), making it conceptually relevant to our examination of trust in technology used for health information seeking. Participants rated 11 items referring to ``AI assistants like ChatGPT or similar tools'' on a 5-point agreement scale (Appendix~\ref{appendix:questions}). Each subscale was computed as an unweighted item mean and later composited into a single $GeneralTrust$ score (mean of all 11 items). We use this composite as a broad measure encompassing both trusting beliefs and intentions toward relying on, sharing information with, and following advice from LLMs. Questions were selected based on the factor loading thresholds in the original paper~\cite{mcknight2002impact}. 
We conducted a confirmatory factor analysis to assess contract validity, with standardized factor loadings ranging from 0.60 to 0.95, meeting or exceeding the practically significant threshold of 0.5~\cite{Hair2024multivariate} (see Appendix~\ref{appendix:measure}).

\begin{itemize}[left=0em]
    \item \scshape{TrustingBelief}: (1) competence: They are effective in providing health information; (2) integrity: I would characterize them as honest; (3) benevolence: They are interested in my well-being, not just their own interests. 
    \item \scshape{TrustingIntention}: (1) I can always rely on them in a tough health situation, (2) I feel that I could count on them to help with a crucial health problem, (3) Faced with a difficult health situation that required me to see a professional for a fee, I would use the hospital/clinic backing these technologies. 
    \item \scshape{TrustingShareInformation}: (1) I would be willing to provide information like my name, address, and phone number to them; (2) I would be willing to share the specifics of my health issue with them. 
    \item \scshape{TrustingFollowAdvice}: (1) I would feel comfortable acting upon the information given to me by them; (2) I would not hesitate to use the information they supplied me; (3) I would confidently act on the advice I was given by them. 
\end{itemize}

\noindent\textbf{\scshape RelativeTrust:} For each of the five task types described in Sec.~\ref{sec:materials}, participants were asked, ``How comfortable would you feel using the following tools for [task]?'' and separately rated (1) LLM and (2) traditional search environment on a 5-point scale. To ensure a shared understanding of the comparison, participants were informed that traditional search retrieves a ranked list of existing web pages and sources that match their search terms, rather than generating conversational responses. An accompanying image illustrated that sources in the search environment can range from institutional resources such as the US Food and Drug Administration (FDA) to social media platforms like YouTube and Reddit. Relative trust was computed as $RelativeTrust = TrustAI - TrustSearch$ for each task, with positive values indicating LLM preference and negative values indicating search preference.

\noindent\textbf{\scshape DecisionConfidence:} After rating their comfort relying on an LLM versus traditional search across the five task types, participants answered ``How confident are you in deciding when and whether to use LLMs for health informational needs''. This captures their perceived competence in making appropriate reliance decisions.

\noindent\textbf{\scshape Individual Differences:} We measured four additional self-reported individual difference variables as potential moderators (mentioned above in Sec.~\ref{sec:procedure}; complete questions provide in Appendix~\ref{appendix:questions}): AI literacy~\cite{jin2025glat}, self-reported AI familiarity, pre-existing AI usage frequency and comfort with using AI for health-related questions, and Need for Cognition scale (NCS-6)~\cite{cacioppo1984efficient, lins2020very}.


\subsection{Analysis}

Prior to hypothesis testing, we examined outcome distributions within each condition. $GeneralTrust$ was approximately symmetric across conditions, although the Shapiro-Wilk test indicated a minor deviation in Condition 4 ($W$ = 0.97, $p$ = .017), and skewness was limited (0.50) and did not indicate a substantial departure from normality~\cite{kim2013statistical}. $DecisionConfidence$ was a single 5-point ordinal item, and five task-specific $RelativeTrust$ scores were non-normal. Lastly, Levene's test confirmed homogeneity of variance across all outcomes (all $p$ > .21).

To test H1, we compared the baseline condition with the combined risk communication conditions. $GeneralTrust$ was tested using an independent-samples $t$-test, and $DecisionConfidence$ using Mann-Whitney $U$ tests because of the ordinal single-item structure. 
To test H2 and H3, we limited the analysis to the four risk communication conditions arranged in a $2\times2$ design (risk context: general versus health-specific; mechanistic explanation: absent versus present; each condition $N$s ranging from 83 to 113). $GeneralTrust$ was analyzed using OLS with risk context, mechanistic explanation, and their interaction as predictors, with HC3-robust standard errors. $DecisionConfidence$ was analyzed using a cumulative-link proportional-odds ordinal logistic regression with the same predictors, while H2b and H3b were evaluated using Wald $z$-tests. The proportional-odds assumption was assessed by comparing coefficients across binary logistic models fitted at each cumulative threshold. Mann-Whitney $U$ tests were additionally used as supplementary robustness checks for each main effect.

For $RelativeTrust$, we analyzed the five health information task types separately. $RelativeTrust$ was operationalized as the difference between ratings of comfort relying on AI and search environments for each task ($LLM - Search$), with positive values indicating greater relative preference for AI. Because these task-specific dimensions were non-normal, Mann-Whitney $U$ tests were used to compare the combined risk communication conditions against the baseline.

Finally, we conducted exploratory moderation analyses to examine whether AI literacy, AI familiarity, AI usage frequency, prior attitudes toward using AI for health-related information, and need for cognition (NFC) moderated the study outcomes. Moderators were mean-centered before moderation analyses. For each moderator-outcome pair, we estimated a separate OLS regression model with HC3-robust standard errors, with 15 interaction tests in total. 

To understand how participants interpreted the risk communication beyond its effects on trust and decision confidence, we analyzed the open-ended reflections collected after each card set. Participants in all four experimental conditions responded after the risk cards, and those in the mechanistic explanation conditions also responded after the mechanism cards. This gave 570 answers (204 following general risk, 178 following health-specific risks, and 188 following mechanism characteristics) from 383 participants. We conducted a reflexive thematic analysis~\cite{braun2006using, braun2019reflecting}. Because card content differed across conditions, we focused on \textit{how} participants engaged with the communication rather than which risks they found concerning. Two authors read all responses and created initial codes inductively. Codes were then organized into themes and iteratively refined through discussions.


\section{Results}

\begin{table}
\sffamily
\centering
\caption{Summary of all test results. The final column indicates whether the hypothesis was supported (\checkmark where  $p <$ .05), borderline ($\cdot$ where  $p <$ .10), or not supported (blank).}
\label{table:result_summary}
\footnotesize
\begin{tabular}{llccc}
\rowcollight\textbf{Hypotheses and exploratory analyses} & \textbf{Statistic}\dag & \textbf{$p$} & \textbf{Effect size $d$ with 95\% CI}\ddag & \\
\toprule
\multicolumn{5}{l}{\textit{H1: Effect of risk communication vs.\ baseline ($N$ = 500)}} \\[4pt]
H1a. Risk comm $\rightarrow$ lower GeneralTrust      & $t = -4.22$ & \textbf{.000} & $-0.45$ $[-0.66, -0.24]$ & \checkmark \\
H1b. Risk comm $\rightarrow$ higher DecisionConf.\   & $U = 20734.0$ & .227    & $-0.14$ $[-0.35, +0.07]$ &  \\[8pt]

\midrule
\multicolumn{5}{l}{\textit{H2: Effect of health-specific context ($N$ = 384)}} \\[4pt]
H2a. Health context $\rightarrow$ lower GeneralTrust   & $F = 5.39$ & \textbf{.021}  & $-0.26$ $[-0.46, -0.05]$  & \checkmark \\
H2b. Health context $\rightarrow$ higher DecisionConf. & $z = 0.40$ & .689 & $+0.06$ $[-0.14, +0.27]$ &  \\[8pt]

\midrule
\multicolumn{5}{l}{\textit{H3: Effect of mechanistic explanation ($N$ = 384)}} \\[4pt]
H3a. Mechanism $\rightarrow$ lower GeneralTrust        & $F = 3.28$ & .071 & $-0.21$ $[-0.41, -0.01]$ & $\cdot$ \\
H3b. Mechanism $\rightarrow$ higher DecisionConf. & $z = -0.29$ & .773 & $-0.06$ $[-0.27, +0.14]$ &  \\[8pt]

\midrule
\multicolumn{5}{l}{\textit{Exploratory: Relative trust by task type ($N$ = 500)}} \\[4pt]
Medical: Risk comm $\rightarrow$ lower relativeAIPref. & $U = 18612.5$ & \textbf{.006}     & $-0.29$ $[-0.50, -0.08]$ & \checkmark  \\
Therapeutic: Risk comm $\rightarrow$ lower relativeAIPref. & $U = 19755.0$ & .055 & $-0.25$ $[-0.44, -0.04]$ & $\cdot$ \\
Companion: Risk comm $\rightarrow$ lower relativeAIPref. & $U = 21586.0$ & .596     & $-0.07$ $[-0.26, +0.14]$ &  \\
Impairment: Risk comm $\rightarrow$ lower relativeAIPref. & $U = 18892.0$ & \textbf{.008}    & $-0.30$ $[-0.51, -0.10]$ & \checkmark  \\
Lifestyle: Risk comm $\rightarrow$ lower relativeAIPref. & $U = 21308.5$ & .441     & $-0.11$ $[-0.30, +0.09]$ &  \\
\bottomrule
\end{tabular}
\\[8pt]
\begin{minipage}{\linewidth}
  \raggedright
  \footnotesize
    \dag\textit{Note.} H1a was tested using an independent-samples $t$-test. For H1b, the primary test was the nonparametric Mann-Whitney $U$ test. H2a and H3a were tested using Type II $F$-tests with $df = (1, 380)$ from factorial OLS models with HC3-robust standard errors. H2b and H3b were tested using Wald $z$-tests from a factorial proportional-odds ordinal logistic regression for the single-item ordinal DecisionConf. outcome. Exploratory RelativeTrust analyses used Mann-Whitney $U$ tests to compare the pooled risk-communication conditions against the baseline within each task type. \\[3pt]
    \ddag\textit{Note.} Reported effect sizes are Cohen's $d$ with 95\% confidence intervals and were estimated separately from the model-based hypothesis tests.
\end{minipage}
\end{table}

\subsection{H1: Effects of Risk Communication}

For the effects of risk communication, we found that \textbf{risk communication significantly reduced people's perceived trust in LLMs}.
Participants who received any structured risk communication ($n$ = 384, $M_{\text{riskcomm}}$ = 2.73, $SD$ = 0.90) reported significantly lower trust than those without explicit exposure ($n$ = 116, $M_{\text{baseline}}$ = 3.13, $SD$ = 0.92), where $t$ = $-$4.22, $p$ < .001, $d$ = $-0.45$, 95\% CI [$-$0.66, $-$0.24] (\ref{table:result_summary}). 

Looking more closely at the trust subscales, Fig.~\ref{fig:trust_line} shows consistent descending patterns in trust with risk communication across all subscales (all $p$ < .001). The shared decline across subscales indicates that risk communication broadly reduced perceived trust in LLMs: participants simultaneously became less likely to believe in LLMs' traits of competence, integrity, and benevolence ($TrustingBelief$), less willing to rely on LLMs in health situations ($TrustingIntention$), more reluctant to share personal health information ($TrustingShareInfo$), and less willing to follow AI advice ($TrustingFollowAdvice$). Overall, we found that people placed greater trust in LLMs' general attributes ($TrustingBelief$) than in the other three action-related subscales, echoing the gaps between general stances on LLMs and situated judgments about when to reply.

On the other hand, \textbf{confidence in reliance decisions remained flat and relatively high across all five conditions}, with no significant differences between participants who received risk communication and those who did not ($U$ = 20734, $p$ = .227). 
The confidence intervals in \ref{fig:confidence_line} were consistently high and overlapped largely across conditions. Notably, the direction of differences (despite non-significant) was opposite to the hypothesis: participants who received risk communication had a slightly lower average decision confidence than those who did not ($M_{\text{riskcomm}}$ = 3.85, SD = 0.91; $M_{\text{baseline}}$ = 3.97, SD = 0.87). 



\begin{center}
\begin{minipage}[t]{0.49\linewidth}
    \centering
    \includegraphics[width=\linewidth]{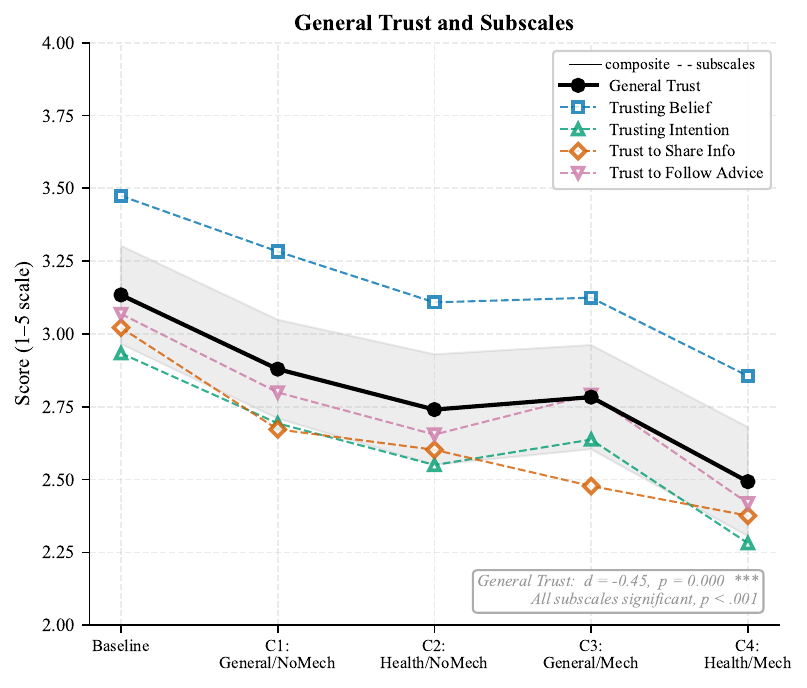}
    \captionof{figure}{Trust in LLM composite and subscale scores by experimental condition (scale [1--5], with 95\% confidence intervals of the composite). Risk communication leads to lower perceived trust in AI for health information compared to no explicit exposure, with significance at the composite effect and subscales (all $p < .001$).}
    \label{fig:trust_line}
    \Description{A line graph shows general trust and four trust subscales across the baseline and four risk communication conditions. Scores are highest in the baseline condition and generally lower across all four risk communication conditions. Among the conditions, C1 (general risks without mechanisms) tends to have higher trust scores, C2 (health risks without mechanisms) and C3 (general risks with mechanisms) show similar intermediate levels, and C4 (health risks with mechanisms) consistently shows the lowest scores. Trusting belief is generally the highest-rated subscale across conditions, while the relative ordering of trusting intention, trust to share information, and trust to follow advice varies across conditions.}
\end{minipage}
\hfill
\begin{minipage}[t]{0.49\linewidth}
    \centering
    \includegraphics[width=\linewidth]{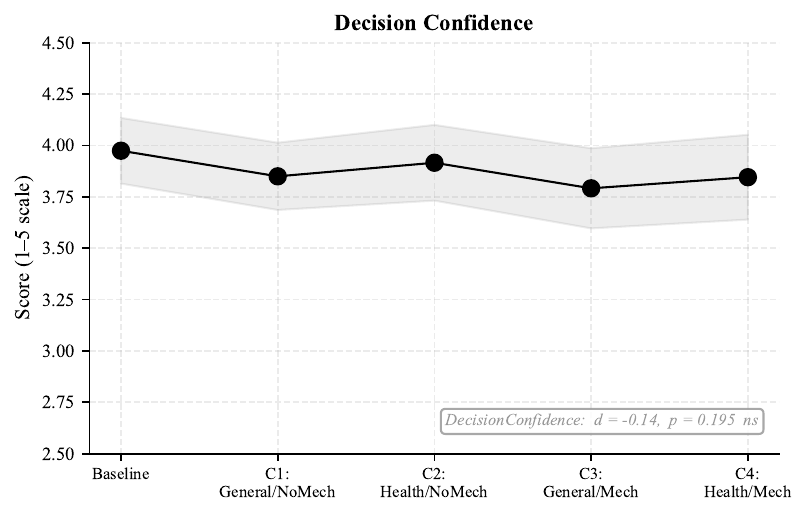}
    \captionof{figure}{Confidence of deciding when to use LLMs for health informational needs by experimental condition (1--5 scale, with 95\% confidence intervals). Scores remain flat across all five conditions ($d = -0.14$, $p = .195$) without significant effects of risk communication on decision self-efficacy about when to rely on LLMs.}
    \label{fig:confidence_line}
    \Description{A line graph shows decision confidence across the five experimental conditions, with mean scores ranging approximately from 3.8 to 4.0.}
\end{minipage}
\end{center}


\subsection{H2: Effects of Health-Specific Context}


For GeneralTrust, the main effect of risk context was statistically significant ($F(1, 380)$ = 5.39, $p$ = .021), indicating that \textbf{risk communication with health-specific focus produced lower general trust than general risk communication}. The Mann-Whitney robustness check confirmed this finding ($p$ = .007). The interaction between context and mechanistic explanation was not significant ($F(1, 380)$ = 0.68, $p$ = .411), indicating that the two factors contributed independently. The incremental reduction can be seen in Figure~\ref{fig:trust_line} through the drop between C1 (General/NoMech) and C2 (Health/NoMech), and again between C3 (General/Mech) and C4 (Health/Mech). For DecisionConfidence, the main effect of context was not significant ($z$ = 0.40, $p$ = .689) with the Mann-Whitney robustness check confirming this null result ($p$ = .528). This means that \textbf{risk communication did not have a significant effect on people's confidence in reliance decisions} (H1b not supported).

\subsection{H3: Effects of Mechanistic Explanation}
\textbf{Mechanistic explanation did not have any significant effect on people’s trust in LLMs or their confidence in reliance decisions.} For GeneralTrust, the main effect of the mechanistic explanation approached but did not reach conventional significance level, with $F(1, 380)$ = 3.28 ($p$ = .071). Therefore, H3a was not supported. The Mann-Whitney robustness check, however, produced a nominally significant result ($p$ = .046), and this directional consistency across both tests suggests a small effect that may warrant examination in a larger sample. 
H3b was also not supported, as the effect of mechanistic explanation on Decision Confidence was non-significant ($z$ = $-$0.29, $p$ = .773), which was further confirmed by the Mann-Whitney robustness check ($p$ = .799).


\subsection{Relative Trust by Health Task Type}

\begin{figure}[h]
    \centering
    \includegraphics[width=0.75\linewidth]{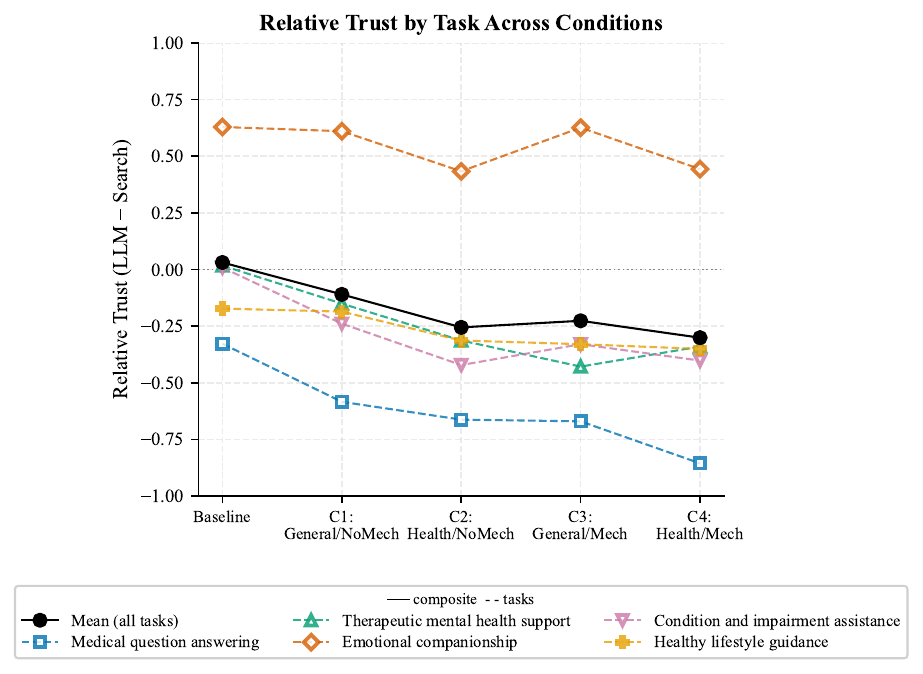}
    \caption{Relative trust (comfort relying on LLM versus traditional search environment) by health task type across experimental conditions. Emotional companionship is the only task that remains consistently positive across all conditions; medical question answering shows the largest and most consistent decline, with both medical question answering and impairment assistance reaching statistical significance (both $p$ < .01).}
    \label{fig:relative_trust}
    \Description{A multi-line graph compares relative trust across five health task types and five experimental conditions. Relative trust is positive for emotional companionship and negative for the other four task types, with medical question answering exhibiting the lowest relative trust across conditions.}
\end{figure}

Fig.~\ref{fig:relative_trust} shows relative trust (i.e., comfort relying on LLMs versus traditional search) by task type across the five conditions. As mentioned in Sec.~\ref{sec:measures}, traditional search refers to keyword-based retrieval systems that return a ranked list of existing web pages and sources, ranging from institutional resources such as the FDA to social media platforms such as YouTube and Reddit, rather than generating answers in a conversational manner. Two health information tasks stand out: emotional companionship was the only task that showed a consistent relative preference for LLMs over navigating traditional search environments across all conditions, while medical question answering showed the lowest relative trust, with a consistent decline across the four different risk communication approaches. 

Across all five task types, participants who received risk communication generally showed lower relative trust in LLMs than those with a generic disclaimer, but the magnitude and significance of this pattern varied across task types. \textbf{Risk communication significantly reduced relative preference for LLMs for two task types: medical question answering and condition \& impairment assistance}. For medical question answering, the Mann-Whitney test indicated a significant difference between risk communication and baseline conditions, $U$ = 18612.5 ($p$ = .006) and $d$ = $-$0.29 with CI [$-$0.50, $-$0.08]. For condition and impairment assistance, a similarly robust pattern was seen, $U$ = 18892.0 ($p$ = .008) and $d$ = $-$0.30 (CI [$-$0.51, $-$0.10]). 
For therapeutic mental health support, although the effect size ($d$ = $-$0.25, CI [$-$0.44, $-$0.04]) is comparable to the two tasks above with significance, the Mann-Whitney test did not reveal significance, $U$ = 19755.0, $p$ = .055. Emotional companionship showed no meaningful difference between conditions ($U$ = 21586.0, $p$ = .596, $d$ = $-$0.07 with CI [$-$0.26, +0.14]). Notably, \textbf{emotional companionship was the only task for which LLMs were the preferred source over search environments, regardless of risk exposure}. Last, healthy lifestyle guidance was similarly non-significant, $U$ = 21308.5, $p$ = .441, $d$ = -0.11 (CI [$-$0.30, +0.09]). 


\subsection{Moderation Effect of AI Literacy and Need for Cognition}

\begin{figure}[tbhp]
\centering
\small
\begin{subfigure}[t]{0.31\columnwidth}
    \includegraphics[width=\linewidth]{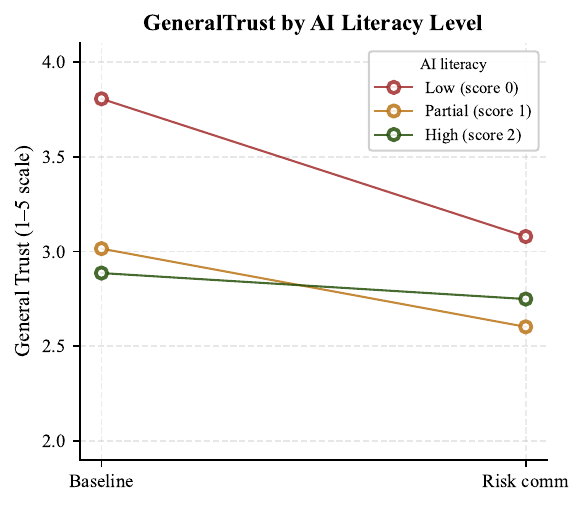}
    \caption{Effects of risk communication on trust by AI literacy ($p$=.033): risk communication brings the trust level of participants with low AI literacy closer to that of participants with high AI literacy.} 
    \label{fig:mod_trust_literacy}
    \Description{A line graph compares general trust between baseline and risk communication conditions across low, partial, and high levels of AI literacy. Trust is substantially higher among participants with low AI literacy at baseline, but the differences across literacy levels narrow following risk communication. Trust decreases substantially among participants with low AI literacy, decreases moderately among those with partial AI literacy, and changes only slightly among those with high AI literacy.}
\end{subfigure}
\hfill
\begin{subfigure}[t]{0.31\columnwidth}
    \includegraphics[width=\linewidth]{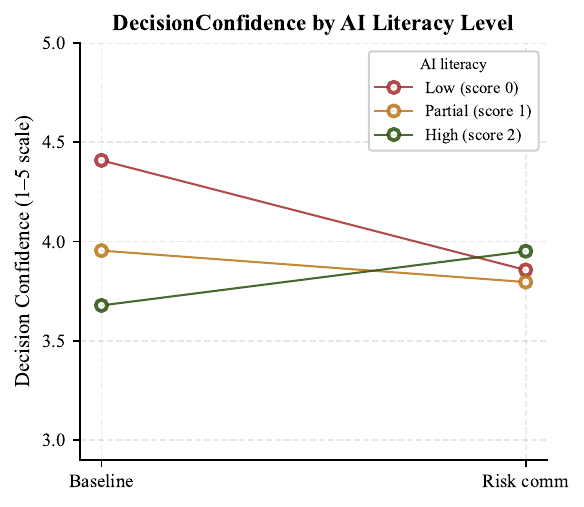}
    \centering
    \caption{Effects of risk communication on decision confidence by AI literacy ($p$=.001): risk communication brings the decision confidence level of participants with lower AI literacy closer to that of participants with high AI literacy.} 
    \label{fig:mod_conf_literacy}
    \Description{A line graph compares decision confidence between baseline and risk communication conditions across low, partial, and high levels of AI literacy. Confidence decreases substantially among participants with low AI literacy, decreases slightly among those with partial AI literacy, and increases among those with high AI literacy. Confidence is highest among participants with low AI literacy at baseline but highest among those with high AI literacy following risk communication.}
\end{subfigure}
\hfill
\begin{subfigure}[t]{0.31\columnwidth}
    \includegraphics[width=\linewidth]{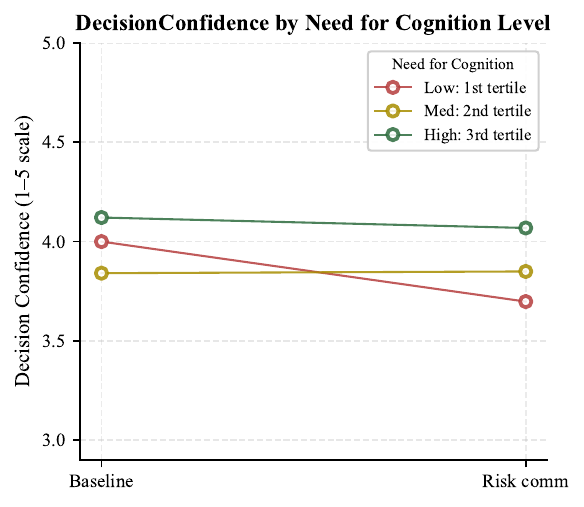}
    \centering
    \caption{Effects of risk communication on decision confidence by need for cognition ($p$=.019): participants who prefer less analytical engagement have decreased confidence in deciding when to trust.} 
    \label{fig:mod_conf_nfc}
    \Description{A line graph compares decision confidence between baseline and risk communication conditions at low, medium, and high levels of need for cognition. Confidence decreases among participants with low need for cognition, remains nearly unchanged among those with medium need for cognition, and decreases slightly among those with high need for cognition.}
\end{subfigure}
\caption{Moderation of risk communication effects on GeneralTrust and DecisionConfidence by AI literacy and Need for Cognition.}
\label{fig:moderation}

\end{figure}

In moderation analyses, three of the 10 moderator-outcome interaction tests were found to be significant: AI literacy moderated both outcomes, and need for cognition moderated DecisionConfidence. All remaining seven interactions were non-significant, showing that AI familiarity, AI usage frequency, and prior AI comfort did not moderate the effect of risk communication on either outcome.

AI literacy moderated the effect of risk communication on both GeneralTrust ($\beta$ = 0.31, p = .033, 95\% CI [0.02, 0.59]) and DecisionConfidence ($\beta$ = 0.42, p = .001, 95\% CI [0.16, 0.67]). The positive coefficient indicates that greater objective knowledge attenuated the reduction in trust and decision confidence. Baseline data (Fig.~\ref{fig:moderation}) shows that participants who knew the least about how LLMs work placed the most trust in LLMs and were the most certain about when to rely on one. Reading these coefficients against the baseline levels, we can see that \textbf{risk communication narrowed the gaps in trust and decision confidence across AI literacy levels}, by bringing participants with lower AI literacy closer to the positions that participants with high AI literacy held. This suggests that risk communication could potentially serve as corrective information to miscalibrated mental models of LLMs. However, contrary to the H1b hypothesis, following risk communication, \textbf{only high-literacy participants showed an increase in confidence of reliance decisions}, suggesting that more AI literacy may be needed to help users to translate risk information into clearer decision-making.

Need for cognition additionally moderated the effect on DecisionConfidence ($\beta$ = 0.23, p = .019, 95\% CI [0.04, 0.42]), with higher need for cognition similarly counteracting the confidence effect. Critically, participants who preferred less analytical engagement showed the largest confidence reduction, suggesting that \textbf{risk communication may reduce confident decision-making among those less disposed to process complex new information}. Together, these two results suggest that the intended effect of risk communication in helping people decide when and whether to rely on LLMs may require people to have the knowledge or disposition to make use of it.


\subsection{How Participants Interpreted Risk Communication}

The open-ended answers revealed four ways participants interpreted the risk communication, including:

\subsubsection{Familiar Warnings Became More Concrete.}
Many participants mentioned that they knew that LLMs can make mistakes, but less about the specific risks. One participant wrote, \textit{``I knew that information was not 100\% accurate but I did not know how convincingly it could be presented as facts''} (P223). Others described their prior knowledge as partial (\textit{``I did at least partially know before but not to this extent''}, P57) or limited (\textit{``what I knew only touches the surface''}, P306). At the same time, participants noted that examples made these risks easier to notice. P204 said \textit{``Some of the responses I thought were fine before it was pointed out what was wrong with them''}. However, examples also sometimes led participants to conclusions beyond the intended message. P113 inferred that an AI referencing a user's age meant \textit{``it's collecting data unknowingly''}, while P331 interpreted a privacy example as showing an AI \textit{``able to pull a real person's medical records''}. The examples made risks more concrete, but participants did not always interpret them as accurately as intended.

\subsubsection{Personal Experience Shaped Credibility.}
Some participants compared the communication with their own experience with LLMs. When the examples differed from what they had seen, they questioned whether they were realistic, or it was because of the model they used or how they wrote prompts. For example, P151 questioned \textit{``This AI gives substantially worse responses than any I have used''} and P115 guessed the difference is because \textit{``I pay \$20 a month for Chat GPT''}. Similar reactions were seen in mechanistic explanations. Participants questioned statements with their experiences, noting that LLMs \textit{``will tell me the source sometimes''} (P288) or \textit{``refer to earlier prompts or messages''} (P18). When an explanation fit prior experience, however, it could make that experience easier to understand: \textit{``This explains a lot about the kinds of results I have gotten''} (P379).

\subsubsection{Caution Without Clear Action.}
A common takeaway noted in participants' answers was to be cautious and verify important information. As P229 explained, \textit{``we can and should verify all important information before acting''}. However, we noticed that participants rarely said what they would verify, and how they would do it.
Some participants pointed to this uncertainty themselves. P276 said that \textit{``(LLM unreliability) seems to make it useless or require a structured way of using it''}, while P50 felt unsure whether \textit{``the rewards outweigh the potential risks in many situations''}. These reflections were consistent with the quantitative finding that risk communication reduced trust without increasing confidence in deciding when to rely on LLMs.

\subsubsection{Mixed Views on Risk and Responsibility.}
Overall, participants did not agree on how much the risks applied to them or who should be responsible for addressing them. First, there was a difference in whether and how participants perceived the risks as relevant to themselves. Some felt they already understood the risks and were more concerned about other users; for example, P377 said \textit{``None of these are of large concern to me personally as I am aware of the risks in my own use. However, I do have concern for others''}. Fewer people recognized that the risks could apply to them: \textit{``Probably erosion of critical thinking happens with me too''} (P207).

Participants also differed in who they thought should address these risks. Some treated the problems as consequences of user behaviors, attributing inappropriate assumptions to \textit{``not telling the AI ahead of time''} (P115) or privacy risks to \textit{``user error by oversharing''} (P266). Others expected AI systems themselves to intervene (\textit{``Every single one of these cases should just produce a warning that says `CONSULT A REAL DOCTOR'''}, P305), or noted a void in regulation (\textit{``no one has yet decided who is liable if the AI is wrong''}, P382).

These different positions were also reflected in participants' reactions to the communication. Some became broadly distrustful, noting that \textit{``It feel like AI can not be trusted''} (P37). Others pushed back on risks involving features they valued (\textit{``Frankly, I wouldn't use my chatGPT at all if it DIDN'T seem so humanlike in communication''}, P69), or even questioned overreaction to risks (\textit{``People can infer a lot just based off a single answer and that is quite unnecessary''}, P15).



\section{Discussion}

Our study offers one of the first controlled experiments on how structured risk communication affects people's trust and confidence in deciding when to rely on LLMs for health information seeking. Results show that risk communication significantly reduced perceived trust, especially when it was contextually relevant. However, this trust calibration did not change how confident they felt in deciding when to rely on them, except for people with high objective knowledge. Task-level effects were similarly uneven: disclosure shifted preferences away from LLMs for medical question answering and impairment \& condition assistance, but participants continued to prefer LLMs over traditional search environments for emotional companionship.
Together, these findings suggest that risk communication can change how people generally evaluate LLMs, particularly when the disclosed risks are related to the context in which they are used. However, recognizing that a system is risky is different from knowing what to do with that information. For general-purpose LLMs, for effective communication to assist risk recognition and situated action, it needs to account for differences in both the use context and the knowledge users bring to the interaction.


\subsection{Risk Communication as a Corrective for Miscalibrated Trust}

At a time when public understanding lags behind technical development and adoption speed~\cite{zhang2024s, belghith2024testing, sula2026rehearsing, steyvers2025large, spatharioti2025effects}, our study provides evidence that risk communication can serve as a corrective mechanism for miscalibrated trust. In our study, before any exposure, participants with the least accurate understanding of how LLMs work reported the highest trust in them and the greatest confidence in deciding when to rely on them. Risk communication significantly reduced trust in LLMs across all four dimensions~\cite{mcknight2002impact} (beliefs about competence, integrity, and benevolence; intention to depend on LLMs; and intention to share information with and follow their advice). Importantly, it brought those with the least accurate understanding of AI closer to the calibration already held by those with the most accurate ones.

Contextual relevance further strengthened this effect. We found that health-specific framing reduced trust beyond what general framing achieved, even though all conditions used health-related example conversations. In other words, showing risks closer to the usage context in the risk message changed how much people trusted the system, despite seeing usage-related examples. This indicates that simply illustrating a general risk with a health example may not make that risk equivalent to communication organized around the consequences and concerns related to the usage case.

Prior practices developed in domains such as health and crisis communication emphasize quantified and causal communication: report absolute rather than relative risks, express uncertainty as natural frequencies, and show exactly which action leads to which consequence~\cite{fischhoff1995risk, dawson2013helping, ancker2025scope, mcclure2007listening}. These techniques sit on the assumptions that the set of possible actions is bounded and that their outcomes can be counted. However, that assumption no longer holds for LLMs. People can ask an LLM about anything, the answers of a given question vary from case to case, and no one can put a reliable number on how often a prompt leads to harm or through what path. What remains available is context. If systems cannot tell users precisely how likely a harm is, they can still show users where it is likely to arise. General framing could leave users to view harm as abstract and distant, while contextually relevant framing give it a recognizable form that the user can imagine encountering.

\subsection{Differentiating Reducing Overtrust from Supporting Reliance Decisions}

Our findings also show that risk communication significantly changed how much people trusted LLMs without changing how confident they felt in deciding when to rely on them, which implies a distinction within AI risk communication. In many domains, risk messages and the recommended action come together: a warning about smoking implies quitting, or a high blood sugar level implies diet changes. Once someone accepts the risk, the actions can be clear. Risk communication for a general-purpose LLM does not work this way. Trust is formed toward the system as a whole, but reliance is decided one task at a time; and as our results indicate, the same system carries different levels of perceived risks when it is used for a medication question, an explanation of medical reports, or a late-night conversation about loneliness. Therefore, telling users that the system is less trustworthy than they assumed may not be enough to guide them in a particular case. We believe that while correcting overtrust is a necessary starting point for more appropriate reliance, it only becomes actionable when users can apply their judgment.

In fact, this dissociation echoes a long-standing observation that risk awareness and actionable guidance are distinct outcomes that do not always co-occur~\cite{gerrard1999effect, fischhoff1995risk}. Threat-related information without a sense of efficacy leads to avoidance rather than adaptive action~\cite{witte1992putting}. In our study, providing mechanistic explanations of how LLMs generate outputs did not close this gap as we hoped. To interpret this by separating participants into AI literacy groups, we found that against these adjusted baseline comparisons, only participants who held accurate prior objective knowledge gained decision confidence. This suggests that risk communication repurposed directly from existing taxonomies requires adaptation to be more widely useful for general populations.

Accordingly, the relevant evaluation criterion for risk communication should extend beyond whether a warning is noticed or whether it lowers trust. If the intended goal is appropriate reliance, interventions should also be evaluated on whether users can translate disclosed risks into better situated judgments: for example, recognizing when verification, comparison with another source, or deferral is warranted.

\subsection{Implications for Designing Risk Communication for General-Purpose LLMs}

\subsubsection{Matching Disclosure to the Task's Risk Profile.}
The task-level findings indicate that risk communication should not treat the use of a general-purpose LLM as a single decision context. Risk communication shifted relative preferences of LLMs over traditional search environments for two tasks: medical question answering and impairment \& condition assistance. Among the other tasks where risk communication had no significant effect, emotional companionship was the only task for which participants consistently preferred LLMs over search environments that include online communities, regardless of risk exposure. 

The differences show that the same disclosure can have different effects depending on what users are trying to accomplish. One possible explanation is that these two tasks are particularly cognitively driven, and their risk surrounds factual accuracy. On the other hand, emotional companionship is relational in nature and valued for its immediacy and lack of judgment~\cite{kim2024mindfuldiary}, qualities that are easier to perceive than the presented risks. Here, the broader implication is that a single communication approach is unlikely to serve use cases with such different risk profiles, where a disclosure effective at reducing overreliance for medical queries may be less applicable to emotionally related use cases, which turns out to be most common among youth~\cite{CSM2025_teen}.

\subsubsection{Accounting for Who Receives the Disclosure.}
The moderation findings suggest that risk communication can backfire when it does not account for who receives it. Disclosure only achieved its intended purpose of increasing decision confidence among participants who already held more accurate knowledge of how LLMs work. Moreover, for users who are not personally motivated to process and integrate with risk communication (i.e., people with lower NFC), disclosure may further introduce confusion without resolution. Therefore, communication needs to be designed to make understanding accessible. Prior work suggests that interventions which interrupt heuristic processing can reduce overreliance even among those who prefer less analytical engagement~\cite{buccinca2021trust}, and that interactive formats work better than text-based ones~\cite{heuer2022comparative}. For AI literacy and appropriate trust specifically, evidence shows that video messages can shape the mental models people attribute to LLMs and affect their reliance on LLM-generated responses~\cite{chen2026presenting}. Future work could explore progressive and in-context forms of disclosure to reduce the mental burden of converting abstract disclosure into judgments, such as short task-relevant cues followed by optional explanations, concrete demonstrations of failure, or prompts that translate a stated limitation into an immediate action like checking professional sources.

\subsubsection{From Risk Communication to AI Literacy.}
Across our results, prior objective AI knowledge appears to condition whether risk information becomes actionable. This does not mean that disclosure should be replaced by broad AI-literacy education. Rather, the two may serve complementary functions. Risk communication can bring a limitation to users’ attention at the moment it becomes relevant, while AI literacy provides the conceptual structure needed to understand what that limitation means and how it should affect subsequent use.

This distinction is especially important for generative AI because its failures tend to be less visible and literacy is not guaranteed from usage. Many prior technologies have more visible failures, such as a broken link that fails to load, or a search that returns no matching records. Generative AI's failures are fluent and presented identically to its successes, with few default cues distinguishing the two~\cite{kim2024m, zhang2024s, yoo2026ai, shi2026siren}. We argue that the less a technology reveals its failures, the more its safe use rests on understanding rather than experience, which is why functional literacy~\cite{long2020ai} with LLMs in some way depends on conceptual literacy in a way it does not for most prior tools. 
 
Therefore, a promising direction is to treat disclosure as an opportunity for situated literacy support rather than as a standalone warning. Recent work on interventions for appropriate reliance points in this direction by pairing disclosure with concrete verification scaffolds~\cite{bo2025rely}. Such practices could also restore some of the distributed judgment that a single synthesized answer removes~\cite{shaw2026thinking}. Accordingly, instead of only stating that an LLM may produce unreliable information, interfaces might connect the risk to a concrete response: why this type of query is difficult for the system, which alternative sources are available, or when escalation to professional experts is needed. Future work can evaluate whether such scaffolds improve not simply trust or confidence, but behavioral indicators of appropriate reliance, such as source verification or external consultation.


\subsection{Limitations and Future Directions}
Several limitations should be considered when interpreting our findings, and some shed light on directions for future research. First, in measuring decision confidence, we intentionally measured this after participants rated their LLM-versus-search preferences among a list of tasks. While this sequencing choice was intended to ground self-reported confidence in concrete task judgments, a single-item measure can have limited reliability and sensitivity compared to a multi-item scale. More broadly, decision confidence captures participants' perceived certainty about when to rely on LLMs rather than whether their reliance decisions were objectively appropriate. Future work can explore whether changes in trust or confidence correspond to improved calibration of actual reliance behavior. 
Second, the risk communication content was drawn from established risk taxonomies, selected for their scholarly acceptance (in the case of the general taxonomy) and contextual fit (in the case of the health-specific taxonomy). However, these taxonomies were not originally developed to communicate risks to general audiences. We chose not to modify the original language to avoid introducing unintentional framing biases, but some content or language may not be fully accessible to people without technical backgrounds. 
Third, the health information scenarios were controlled to minimize confounding, which necessarily traded some ecological validity for experimental rigor. Participants' responses to scripted conversation vignettes may not fully reflect how they would ask or respond to real LLM interactions where outputs are unpredictable and tailored.
Then, from a long-term perspective, participants in our experiment were exposed to risk communication only once. Whether such effects persist over repeated AI use remains an open question, and future work with longitudinal designs can examine the long-term effects: whether initial disclosure effects last or require reinforcement over repeated AI interactions. 
Finally, although randomization checks confirmed that the five conditions were comparable, the Prolific participant pool can skew younger and more educated than the broader U.S. population. This limits the generalizability of findings to groups with lower digital or health literacy, populations who may be most at risk of miscalibrated AI reliance and who arguably could benefit most from effective risk communication.


\section{Conclusion}

This study examined whether structured risk communication can support a more appropriate trust level and confidence in deciding when to rely on LLMs for health information seeking. In a pre-registered experiment with 500 participants, risk communication significantly reduced trust in LLMs, particularly when risks were health-related. But it did not improve confidence in deciding when to rely on them. Disclosure's effects also varied by task and prior AI knowledge: risk communication shifted preferences between LLMs and traditional search environments for some health tasks, while participants with lower AI literacy tended to show larger trust reductions without corresponding gains in decision confidence. Together, these findings show that correcting overly positive perceptions of AI is not the same as enabling informed reliance. Effective AI risk communication should therefore move beyond generic disclosure toward context-sensitive and accessible approaches that help users translate knowledge of AI risks into actionable judgments about when and how to rely on these systems.


\begin{acks}
Part of this research was supported with a grant from the US Centers for Disease Control and Prevention.
\end{acks}

\section*{Conflict of Interest}
De Choudhury serves on the Expert Council on Well-Being and AI at OpenAI.



\bibliographystyle{ACM-Reference-Format}
\bibliography{references}

\appendix

\section{Survey questions}
\label{appendix:questions}

\textbf{\scshape General Attitudes and Preferences}

\noindent When you seek health-related information, what are your main motivations? (Select all that apply)\\
$\square$ Asking specific questions about illnesses, treatments, test results, or medical terms to better understand health information.\\
$\square$ Seeking therapeutic guidance and coping-oriented assistance for people experiencing stress, anxiety, low mood, or other emotional difficulties.\\
$\square$ Seeking a sense of being heard and accompanied through conversation, especially when someone feels lonely or just wants to talk.\\
$\square$ Seeking help to manage daily activities when someone has a condition or impairment (e.g., chronic pain, mobility limits, vision/hearing changes), such as practical tips and accommodations.\\
$\square$ Seeking guidance on lifestyle habits like eating, physical activity, sleep, and daily routines to support overall health.\\
$\square$ Other (Please specify)\\

\noindent Which of the following sources do you typically use when seeking health-related information? \\
(Scale: Never, Rarely, Sometimes, Often, Always)\\
$\square$ Traditional search engines (e.g. Google) \textit{If you heavily rely on AI-embedded search engines or functions (such as AI Overviews in Google Search, Bing Chat, Perplexity), please select ``AI tools or chatbots''}\\
$\square$ Expert advice (e.g., nurses, doctors, specialists)\\
$\square$ Someone I personally know, like family or friends\\
$\square$ AI tools or chatbots (e.g., Bing Chat, Gemini, ChatGPT, Perplexity)\\

\noindent Within traditional search engines... which of the following sources do you typically use? \\
(Scale: Never, Rarely, Sometimes, Often, Always)\\
$\square$ Online communities and forums (e.g., Reddit groups, Facebook groups, health forums)\\
$\square$ Social media (e.g., YouTube, Instagram, TikTok)\\
$\square$ Expert-written websites (e.g., Mayo Clinic, Psychology Today, WebMD)\\
$\square$ (Inter)Governmental resources (e.g., FDA, WHO)\\

\noindent How confident are you in finding reliable and accurate health information through your usual sources?\\
$\square$ Not confident at all\\
$\square$ Slightly confident\\
$\square$ Somewhat confident\\
$\square$ Fairly confident\\
$\square$ Very confident\\

\noindent How often do you feel that the information you find through your usual health sources fully meets your needs or answers your questions?\\
$\square$ Never\\
$\square$ Rarely\\
$\square$ Sometimes\\
$\square$ Often\\
$\square$ Always\\

\noindent \textbf{\scshape AI Knowledge and Usage}

\noindent How familiar are you with language AI and large language models (LLMs) such as ChatGPT, Gemini, and Bing Chat?\\
$\square$ Not familiar at all: I don't have a basic idea of what they are\\
$\square$ Slightly familiar: I have some idea of what they are\\
$\square$ Moderately familiar: I know what they are and their basic functions and uses\\
$\square$ Very familiar: I have a solid understanding of their capabilities, limitations, and applications\\
$\square$ Extremely familiar: I have expert-level knowledge, including their technical aspects, development, and how they work\\

\noindent Do you use language AI and LLMs such as ChatGPT, Gemini, and Bing Chat?\\
$\square$ Never used\\
$\square$ Rarely, use it only occasionally\\
$\square$ Sometimes, but not on a regular basis\\
$\square$ Often, as part of my routine\\
$\square$ Always, use it consistently whenever applicable\\

\noindent How often do you use language AI and LLMs (including heavily relying on AI Overview of Google Search) to find information or answers to questions?\\
$\square$ Never\\
$\square$ Rarely: I use them only in specific situations\\
$\square$ Sometimes: I use them from time to time\\
$\square$ Often: I use them regularly\\
$\square$ Always: I consider them a primary source for finding information\\

\noindent How do you feel about using AI tools to help answer health-related questions?\\
$\square$ Very uncomfortable\\
$\square$ Uncomfortable\\
$\square$ Neutral\\
$\square$ Comfortable\\
$\square$ Very comfortable\\

\noindent \textit{The questions below are about AI — just pick whatever feels most accurate to you, no need to look anything up. There are no trick questions! We're just curious what you think.}\\

\noindent Which of the following statements best describes a Large Language Model (LLM)?\\
$\square$ It generates text by analyzing and summarizing large volumes of web content.\\
$\square$ It generates text by predicting the next word based on the context of previous words.\\
$\square$ It generates text by translating input text into multiple languages simultaneously.\\
$\square$ It generates text by using pre-defined templates and filling in the blanks.\\

\noindent When using a Large Language Model (LLM) to gather information for an assignment, how should you approach the information it provides?\\
$\square$ The LLM's answers are always more trustworthy than any information you will find on the internet, so you can use them without further verification.\\
$\square$ The LLM's answers are generally more trustworthy than internet sources, but you should still verify the information with other reliable sources.\\
$\square$ The LLM's answers are not necessarily more trustworthy than internet sources, and you should cross-check the information with other credible references.\\
$\square$ The LLM's answers are less trustworthy than internet sources because it relies on outdated information.\\

\noindent \textbf{\scshape Trust and Attitudes}

\noindent For language AI and large language models (LLMs)... I think: \\
(Scale: 1 - Strongly Disagree to 5 - Strongly Agree)
\begin{enumerate}[left=1em]
    \item They are competent and effective in providing health information
    \item I would characterize them as honest
    \item They are interested in my well-being, not just their own interests
    \item I can always rely on them in a tough health situation
    \item I feel that I could count on them to help with a crucial health problem
    \item Faced with a difficult health situation that required me to see a professional for a fee, I would use the hospital/clinic backing these technologies
    \item I would be willing to provide information like my name, address, and phone number to them
    \item I would be willing to share the specifics of my health issue with them
    \item I would feel comfortable acting upon the information given to me by them
    \item I would not hesitate to use the information they supplied me
    \item I would confidently act on the health advice I was given by them
\end{enumerate}

\noindent How comfortable would you feel using \texttt{[LLM vs Traditional search engine]} for: \\
(Scale: 1 - Very uncomfortable to 5 - Very comfortable)\\
\vspace{-1em}
\begin{itemize}[left=1em]
    \item (...) Medical Question Answering?
    \item (...) Therapeutic Mental Health Support?
    \item (...) Emotional Companionship?
    \item (...) Impairment and Condition Assistance?
    \item (...) Healthy Lifestyle Guidance?
    \item (...) Select Very Uncomfortable?
\end{itemize}

\noindent How confident are you in deciding when and whether to use LLMs for health informational needs?\\
$\square$ Very unconfident\\
$\square$ Unconfident\\
$\square$ Neutral\\
$\square$ Confident\\
$\square$ Very confident\\

\noindent \textbf{\scshape Demographics and Preferences}

\noindent What is your age?\\
$\square$ 18-24\\
$\square$ 25-34\\
$\square$ 35-44\\
$\square$ 45-54\\
$\square$ 55-64\\
$\square$ 65 or older\\

\noindent What is your gender?\\
$\square$ Female\\
$\square$ Male\\
$\square$ Non-binary\\
$\square$ Prefer not to say\\

\noindent Which best describes your ethnicity?\\
$\square$ American Indian or Alaska Native\\
$\square$ Asian\\
$\square$ Black or African American\\
$\square$ Hispanic or Latino or Spanish Origin\\
$\square$ Native Hawaiian or Pacific Islander\\
$\square$ White\\
$\square$ Two or more races\\
$\square$ Other (please specify)\\
$\square$ Prefer not to say\\

\noindent What is the highest level of school you have completed?\\
$\square$ Some high school or less\\
$\square$ High school graduate (Diploma or GED)\\
$\square$ Some college but no degree\\
$\square$ Associate degree (2-year degree)\\
$\square$ Bachelor's degree (4-year degree)\\
$\square$ Advanced degree (Master's, Professional, or Doctorate degree)\\
$\square$ Prefer not to say\\

\noindent Please indicate whether or not the statement is characteristic of your or what you believe:\\
(Scale: 1 - Extremely Uncharacteristic to 5 - Extremely Characteristic)\\
\begin{itemize}[left=1em]
    \item I would prefer complex to simple problems
    \item I like to have the responsibility of handling a situation that requires a lot of thinking
    \item Thinking is not my idea of fun
    \item I would rather do something that requires little thought than something that is sure to challenge my thinking abilities
    \item I really enjoy a task that involves coming up with new solutions to problems
    \item I would prefer a task that is intellectual, difficult, and important to one that is somewhat important but does not require much thought
\end{itemize}

\section{Example Consumer Health Task Scenarios}
\label{appendix:scenarios}

\textbf{\scshape Medical Question Answering}\\
\textit{(Asking specific questions about illnesses, treatments, test results, or medical terms to better understand health information.)}

\noindent A person your age recently had a routine blood test as part of a general checkup. The results came back with a note flagging that their blood sugar level is slightly above normal, and the doctor's summary mentions the words "pre-diabetes." Their next appointment is in two weeks. They turned to an AI to understand what this means.

\begin{quote}
\textbf{User:} ``My blood test came back and it says my blood sugar is slightly high. The doctor mentioned pre-diabetes. What does that mean exactly?''

\medskip
\textbf{AI:} ``Pre-diabetes means your blood sugar levels are higher than normal but not yet high enough to be diagnosed as type 2 diabetes. It's essentially an early warning sign that your body is having difficulty managing blood sugar effectively\\
At this stage, it doesn't mean you will develop diabetes — many people with pre-diabetes never progress to type 2, especially with some lifestyle changes. The most commonly recommended steps involve adjustments to diet and physical activity, though what's most appropriate depends on a person's individual circumstances.\\
It's worth having a detailed conversation with your doctor at your next appointment about what your specific result means and what, if anything, they'd recommend for you.''
\end{quote}

\noindent\textbf{\scshape Therapeutic Mental Health Support}\\
\textit{(Providing guidance and coping-oriented assistance for people experiencing stress, anxiety, low mood, or other emotional difficulties.)}

\noindent A person your age has been struggling with persistent low mood for the past few months. They haven't spoken to a therapist, partly because they're not sure their problems are "serious enough" to warrant it. One evening they decided to talk it through with an AI instead.

\begin{quote}
\textbf{User:} ``I've been feeling pretty low for a few months now. Nothing dramatic has happened, I just feel kind of flat and unmotivated most of the time. I haven't talked to anyone about it. I'm not sure if it's even worth worrying about.''

\medskip
\textbf{AI:} ``It sounds like you've been carrying this quietly for a while — and the fact that you're not sure it's worth worrying about is itself worth paying attention to. Feeling flat and unmotivated for months, even without a clear cause, is something that deserves to be taken seriously, not dismissed.\\
Sometimes the absence of a dramatic reason makes it harder to give yourself permission to feel the way you do. But low mood doesn't need a specific trigger to be real or to matter.\\
Can I ask — when you say flat and unmotivated, is it more that things feel joyless, or more that it feels hard to get started with things, or both?
''

\end{quote}

\noindent\textbf{\scshape Emotional Companionship}\\
\textit{(Providing a sense of being heard and accompanied through conversation, especially when someone feels lonely or just wants to talk.)}

\noindent A person your age moved to a new city six months ago for a job opportunity. The move made sense at the time, but they haven't found it easy to build a social life and most evenings are spent alone. They feel the absence of familiar people and places. One evening they started talking to an AI just to have some company.

\begin{quote}
\textbf{User:} ``I moved to a new city six months ago and I still feel like I haven't really settled. I don't have many people here yet and I miss home. I'm fine, I just feel a bit lonely I guess.''

\medskip
\textbf{AI:} ``Six months can feel like it should be enough time to feel at home somewhere, but it often isn't — especially when you're building a social life from scratch in a place where you don't have shared history with anyone yet.\\
Missing home isn't just about the place — it's usually about the ease of being around people who already know you, where you don't have to work at connection. That kind of comfort takes time to rebuild somewhere new, and until it exists, even a city full of people can feel isolating.
What do you miss most about where you were before?''

\end{quote}

\noindent\textbf{\scshape Impairment and Condition Assistance}\\
\textit{(Helping to manage daily activities when someone has a condition or impairment (e.g., chronic pain, mobility limits, vision/hearing changes), such as practical tips and accommodations.)}

\noindent A person your age has been dealing with persistent knee pain for the past year, diagnosed as early-stage osteoarthritis. Their doctor has told them to stay active but avoid high-impact activities. They're finding that ordinary daily tasks — climbing stairs, getting up from a chair, walking for longer periods — have become uncomfortable and are starting to limit what they do. They turned to an AI for practical advice.

\begin{quote}
\textbf{User:} ``I have knee pain from early osteoarthritis and everyday things like stairs and getting up from chairs have become really uncomfortable. My doctor said to stay active but avoid high impact. What can I do to manage day to day?''

\medskip
\textbf{AI:} ``Knee osteoarthritis makes a lot of ordinary movements uncomfortable, and staying active while avoiding high impact is the right general direction — movement helps maintain the joint and the muscles around it.\\
A few everyday adjustments tend to help: leading with your stronger leg on stairs, using your arms to push up from chairs, and avoiding seats that are very low all reduce the load on the affected knee.\\
The specifics of what helps most really depend on which activities are giving you the most trouble day to day. What's been hardest for you lately?''

\end{quote}

\noindent\textbf{\scshape Healthy Lifestyle Guidance}\\
\textit{(Providing guidance on habits like eating, physical activity, sleep, and daily routines to support overall health.)}

\noindent A person your age has been trying to eat more healthily for a while but is finding it hard to make consistent changes. They eat reasonably well during the day but evenings are harder — they tend to reach for convenient, less nutritious options when tired after work. They're not looking for a strict diet plan, just practical ways to build better habits. They turned to an AI for guidance.

\begin{quote}
\textbf{User:} ``I want to eat more healthily but I struggle to keep it up, especially in the evenings when I'm tired. I don't want a strict diet — just some practical ways to build better habits. Where do I start?''

\medskip
\textbf{AI:} ``The evening pattern you're describing — doing well during the day but defaulting to less nutritious options when tired — is very common. Tiredness reduces the mental energy available for decisions, so convenience tends to win.\\
A few things that tend to help: having easier, reasonably nutritious options already available in the evening, and eating enough during the day so you're not arriving home genuinely hungry on top of tired.\\
Small, specific changes tend to stick better than broad intentions — picking one habit to adjust first usually works better than overhauling everything at once. What does your typical evening look like?''
\end{quote}

\section{Construct Validity: Exploratory Factor Analysis}
\label{appendix:measure}

\begin{table}[tbhp]
    \centering
    \begin{tabular}{lcc}
    Subscale&Item& Factor loading\\
    \toprule
    $TrustBelief$ & TB1&  .84 \\
    &TB2&  .80 \\
    &TB3&  .74 \\
    $TrustIntention$ &TI1&  .90 \\
    &TI2&  .90 \\
    &TI3&  .60 \\
    $IntentionShareInfo$ &SI1&  .68 \\
    &SI2&  .81 \\
    $IntentionFollowAdvice$ &FA1& .89 \\
    &FA2& .87 \\
    &FA3& .95 \\
    \end{tabular}
    \caption{Results of exploratory Factor Analysis}
\end{table}

Internal consistency was also assessed prior to analysis, and it was excellent for the final composite score ($\alpha$ = .942) and good to excellent for all four subscales (Trusting Belief: $\alpha$ = .831; Trusting Intention: $\alpha$ = .825; Intention to Share Information: $\alpha$ = .708; Intention to Follow Advice: $\alpha$ = .929).

\end{document}
\endinput